\documentclass{article}

\usepackage{times}
\usepackage{subfigure}

\usepackage{natbib}

\usepackage{algorithm}
\usepackage{algorithmic}

\usepackage{amsmath}
\usepackage{amssymb}
\usepackage{url}
\usepackage{graphicx,pstricks}
\usepackage{graphics}
\usepackage{epstopdf}
\usepackage{moreverb}
\usepackage{epsfig}
\usepackage{subfigure}
\usepackage{wrapfig}
\usepackage{hangcaption}
\usepackage{txfonts}
\usepackage{palatino}
\newtheorem{theorem}{Theorem}

\newtheorem{corollary}{Corollary}

\newtheorem{definition}{Definition}

\usepackage[accepted,nohyperref]{icml2016_arxiv}

\icmltitlerunning{Exact Weighted Minwise Hashing in Constant Time}

\begin{document}

\twocolumn[
\icmltitle{Exact Weighted Minwise Hashing in Constant Time}
\icmlauthor{Anshumali Shrivastava}{anshumali@rice.edu}
\icmladdress{Department of Computer Science,
            Rice University, Houston, TX, USA}
\icmlkeywords{Weighted Minwise Hashing, Approximation Algorithms, Big-Data, Large Scale Learning}

\vskip 0.3in
]

\begin{abstract}
Weighted minwise hashing (WMH) is one of the fundamental subroutine, required by many celebrated approximation algorithms, commonly adopted in industrial practice for large scale-search and learning. The resource bottleneck of the algorithms is the computation of multiple (typically a few hundreds to thousands) independent hashes of the data. The fastest hashing algorithm is by Ioffe~\cite{Proc:Ioffe_ICDM10}, which requires one pass over the entire data vector, $O(d)$ ($d$ is the number of non-zeros), for computing one hash. However, the requirement of multiple hashes demands hundreds or thousands passes over the data. This is very costly for modern massive dataset.

In this work, we break this expensive barrier and show an expected constant amortized time algorithm which computes $k$ independent and unbiased WMH in time $O(k)$ instead of $O(dk)$ required by Ioffe's method. Moreover, our proposal only needs a few bits (5 - 9 bits) of storage per hash value compared to around $64$ bits required by the state-of-art-methodologies. Experimental evaluations, on real datasets, show that for computing 500 WMH, our proposal can be 60000x faster than the Ioffe's method without losing any accuracy. Our method is also around 100x faster than approximate heuristics capitalizing on the efficient ``densified" one permutation hashing schemes~\cite{Proc:OneHashLSH_ICML14}. Given the simplicity of our approach and its significant advantages, we hope that it will replace existing implementations in practice.
\end{abstract}

\section{Introduction and Motivation}

(Weighted) Minwise hashing (or Minwise Sampling) \cite{Proc:Broder,Proc:Broder_WWW97,Report:Manasse_00} is the most popular and successful randomized hashing technique~\cite{Proc:Shrivastava_AISTATS14}, commonly deployed in commercial big-data systems~\cite{Article:Henzinger04,Proc:Henzinger_06} for large scale search and learning. Minwise hashing was originally developed for fast similarity estimation. However, quickly it was found that the same sampling (or hashing) scheme made several applications over the web, such as, near-duplicate detection~\cite{Proc:Broder_FUN98,Proc:Henzinger_06}, all-pairs similarity~\cite{Proc:Bayardo_WWW07}, record linkage~\cite{Proc:Koudas_SIGMOD06}, temporal correlations~\cite{Proc:Chien_WWW05}, etc., efficient by drastically reducing their computational and memory requirements.

The idea underlying Minwise hashing is also the key breakthrough that led to the theory of Locality Sensitive Hashing (LSH)~\cite{Proc:Indyk_STOC98}. In particular, minwise sampling is a known LSH for the Jaccard similarity~\cite{Book:Raj_Ullman}. Given two positive vectors $x, \ y \in \mathbb{R^D}$, $x, \ y > 0$, the (generalized) Jaccard similarity is defined as
\begin{equation}\label{eq:WJaccardDef}
  \mathbb{J}(x,y) = \frac{\sum_{i=1}^D\min\{x_i,y_i\}}{\sum_{i=1}^D\max\{x_i,y_i\}}.
\end{equation}
$\mathbb{J}(x,y)$ is a frequently used measure for comparing web-documents~\cite{Proc:Broder}, histograms (specially images~\cite{Proc:Ioffe_ICDM10}), gene sequences~\cite{Proc:Rasheed_SIAM13}, etc. Recently, it was shown to be a very effective kernel for large-scale non-linear learning~\cite{Proc:Li_KDD15}. Furthermore, it was found that WMH leads to the best-known LSH for $L_1$ distance~\cite{Proc:Ioffe_ICDM10}, commonly used in computer vision, improving over~\cite{Proc:Datar_SCG04}.

Weighted Minwise Hashing (WMH) (or Minwise Sampling) generates randomized hash (or fingerprint) $h(x)$, of the given data vector $x \ge 0$,  such that for any pair of vectors $x$ and $y$, the probability of hash collision (or agreement of hash values) is given by,
\begin{equation}\label{eq:CollProb}
Pr(h(x) = h(y)) = \frac{\sum \min\{x_i,y_i\}}{\sum \max\{x_i,y_i\}} = \mathbb{J}(x,y).\end{equation} A notable special case is when $x$ and $y$ are binary (or sets), i.e. $x_i, y_i \in \{0,\ 1\}^D$ . For this case, the similarity measure boils down to $\mathbb{J}(x,y) = \frac{\sum \min\{x_i,y_i\}}{\sum \max\{x_i,y_i\}} = \frac{|x \cap y|}{|x \cup y|}$.

Being able to generate a randomized signature, $h(x)$, satisfying Equation~\ref{eq:CollProb} is the key breakthrough behind some of the best-known approximations algorithms for metric labelling~\cite{Proc:Kleinberg_FOCS99}, metric embedding~\cite{Proc:Charikar}, mechanism design and differential privacy~\cite{Artice:Dwork_14}.

A typical requirement for algorithms relying on minwise hashing is to generate, some large enough, $k$ independent Minwise hashes (or fingerprints) of the data vector $x$, i.e. compute $h_i(x) \ i \in \{1,\ 2,...,\ k\}$ repeatedly with independent randomization. These independent hashes can then be used for a variety of data mining tasks such as cheap similarity estimation, indexing for sublinear-search, kernel features for large scale learning, etc. The bottleneck step in all these applications is the costly computation of the multiple hashes, requiring multiple passes over the data. The number of required hashes typically ranges from few hundreds to several thousand~\cite{Proc:OneHashLSH_ICML14,Proc:Shrivastava_UAI14}. For example, the number of hashes required by the famous LSH algorithm is $O(n^{\rho})$ which grows with the size of the data. \cite{Proc:Li_KDD15} showed the necessity of around 4000 hashes per data vector in large-scale learning with $\mathbb{J}(x,y)$ as kernel, making hash generation the most costly step.

Owing to the significance of WMH and its impact in practice, there is a series of work over the last decade trying to reduce the costly computation cost associated with hash generation~\cite{Report:Haeupler_arXiv14}.

The first groundbreaking work on Minwise hashing~\cite{Proc:Broder} computed hashes $h(x)$ only for unweighted sets $x$ (or  binary vectors), i.e. when the vector components $x_i$s can only take values 0 and 1. Later it was realized that vectors with positive integer weights, which are equivalent to weighted sets, can be reduced to unweighted set by replicating elements in proportion to their weights~\cite{Proc:Gollapudi_06CIKM,Report:Haeupler_arXiv14}. This scheme was very expensive due to blowup in the number of elements caused by replications. Also, it cannot handle real weights. In~\cite{Report:Haeupler_arXiv14}, the authors showed few approximate solutions based on ideas developed in~\cite{Proc:Gollapudi_06CIKM,Proc:Kleinberg_FOCS99} to reduce these replications.

Later \cite{Report:Manasse_00}, introduced the concept of consistent weighted sampling (CWS), which focuses on sampling directly from some well-tailored distribution to avoid any replication. This method, unlike previous ones, could handle real weights exactly. Going a step further, Ioffe~\cite{Proc:Ioffe_ICDM10} was able to compute the exact distribution of minwise sampling leading to a scheme with worst case $O(d)$, where $d$ is the number of non-zeros. This is the fastest known exact weighted minwise sampling scheme so far. Since this will be our main baseline, we review it in Section~\ref{sec:review}.

$O(dk)$ for computing $k$ independent hashes is very expensive for modern massive datasets, especially when $k$ with ranges up to thousands. Recently, there was a big success for the binary case, where using the novel idea of ``Densification"~\cite{Proc:OneHashLSH_ICML14,Proc:Shrivastava_UAI14} the computation time for unweighted minwise was brought down to $O(d+k)$. This resulted into over 100-1000 fold improvement. However, this speedup was limited only to binary vectors. Moreover, the samples were not completely independent.

Capitalizing on recent advances for fast unweighted minwise hashing, \cite{Report:Haeupler_arXiv14} exploited the old idea of replication to convert weighted sets into unweighted sets. To deal with non-integer weights, the method samples the coordinates with probabilities proportional to leftover weights. The overall process converts the weighted minwise sampling to an unweighted problem, however, at a cost of incurring some bias (see Algorithm~\ref{alg:UnWted}).  This scheme is faster than Ioffe's scheme but, unlike other prior works on CWS, it is not exact and leads to biased and correlated samples. Moreover, it requires strong independence~\cite{Article:Indyk2001} leading to poor estimation.

All these lines of work lead to a natural question: does there exist an unbiased and independent WMH scheme with same property as Ioffe's hashes but significantly faster than all existing methodologies? We answer this question positively.

\subsection{Our Contributions:}
\begin{enumerate}
  \item We provide an unbiased weighted minwise hashing scheme, where each sampling scheme takes constant time in expectation. This improves upon the best-known scheme in the literature by Ioffe~\cite{Proc:Ioffe_ICDM10} which takes $O(d)$, where $d$ is the number of non-zeros in the data vector.
  \item Our hashing scheme requires much fewer bits usually (5-9) bits instead of 64 bits (or higher) required by existing schemes, leading to around 8x savings in space.
  \item We derive our scheme from elementary first principles. Our scheme is simple and it only requires access to simple uniform random number generator, instead of costly sampling needed by other methods. The hashing procedure is different from traditional schemes and could be of independent interest in itself.
  \item Experimental evaluations on real image histograms show more than 60000x speedup over the best known exact scheme and around 100x times faster than biased approximate schemes based on the recent idea of fast minwise hashing.
  \item Weighted Minwise sampling is a fundamental subroutine in many celebrated approximation algorithms. Some of the immediate consequences of our proposal are as follows:
  \begin{itemize}
    \item We obtain an algorithmic improvement, over the query time of LSH based algorithm, for $L_1$ distance and Jaccard Similarity search. In particular, we reduce the worst-case query time of $(K,L)$ parameterized LSH algorithm~\cite{Proc:Andoni_FOCS06} from O(dKL) to mere $O(KL + dL)$.
    \item We reduce the kernel feature~\cite{Proc:Rahimi_NIPS07} computation time with min-max kernels~\cite{Proc:Li_KDD15}.
    \item We reduce the sketching time for fast estimation of a variety of measures, including $L_1$  and earth mover distance~\cite{Proc:Kleinberg_FOCS99,Proc:Charikar}.
  \end{itemize}
\end{enumerate}

\section{Review: Ioffe's Algorithm and Fast Unweighted Minwise Hashing}
\label{sec:review}

\begin{algorithm}[h!]
\caption{Ioffe's CWS~\cite{Proc:Ioffe_ICDM10}}
\label{alg:Ioffe}
\begin{algorithmic}
\INPUT Vector $x$, random seed[][]

\vspace{0.1in}

\FOR{$i=1$ {\bfseries to} $k$}
\FOR{Iterate over $x_j$ s.t $x_j > 0$ }
\STATE $randomseed = seed[i][j]$;
\STATE Sample $r_{i,j}, \ c_{i,j} \sim Gamma(2,1)$.
\STATE Sample $\beta_{i,j}  \sim  Uniform(0,1)$
\STATE $t_{j} = \bigg\lfloor\frac{\log{x_j}}{r_{i,j}} + \beta_{i,j}\bigg\rfloor$
\STATE $y_j = exp(r_{i,j}(t_{j} - \beta_{i,j}))$
\STATE $z_j = y_j*exp(r_{i,j})$
\STATE $a_j = c_{i,j}/z_j$
\ENDFOR
 \STATE $k^* = \arg\min_j {a_j}$
 \STATE $HashPairs[i] = (k^*,t_{{k^*}})$
\ENDFOR
\STATE {\bf RETURN} HashPairs[]
\end{algorithmic}
\end{algorithm}

We briefly review the state-of-the-art methodologies for Weighted Minwise Hashing (WMH). Since WMH is only defined for weighted sets,  our vectors under consideration will always be positive, i.e. every $x_i \ge 0$.  $D$ will denote the dimensionality of the data, and we will use $d$ to denote the number (or the average) of non-zeros of the vector(s) under consideration.

The fastest known scheme for exact weighted minwise hashing is based on an elegant derivation of the exact sampling process for ``Consistent Weighted Sampling" (CWS) due to Ioffe~\cite{Proc:Ioffe_ICDM10}. This made computation of a single weighted minwise hash $h(x)$ possible in $O(d)$. Although this is still slow for modern massive datasets, it is a good improvement over the prior works which rely on expensive rejection based sampling to sample from the CWS distribution~\cite{Report:Manasse_00}. Ioffe's overall algorithm, samples from some smartly tailored distribution to achieve the required Equation~\ref{eq:CollProb}. The overall procedure is summarized in Algorithm~\ref{alg:Ioffe}.

$O(d)$ for a single hash computation is quite expensive. Even the unweighted case of minwise hashing had complexity $O(d)$ per hashes, until recently. \cite{Proc:OneHashLSH_ICML14,Proc:Shrivastava_UAI14} showed a new one permutation based scheme for generating $k$ near-independent unweighted minwise hashes in $O(d + k)$ breaking the old $O(dk)$ barrier. However, this improvement does not directly extend to the weighted case.

It was known that with some bias, weighted minwise sampling can be reduced to an unweighted minwise sampling using the idea of replicating weights in proportion to their probabilities~\cite{Proc:Gollapudi_06CIKM,Proc:Kleinberg_FOCS99}. Algorithm~\ref{alg:UnWted} describes such a procedure.  A reasonable idea is then to use the fast unweighted hashing scheme, on the top of this biased approximation. The inside for-loop in Algorithm~\ref{alg:UnWted} blows up the number of non-zeros in the returned unweighted set. This makes the process slower and dependent on the magnitude of weights. Moreover, unweighted sampling requires very costly random permutations for good accuracy~\cite{Patrascu:2010:KRL:1880918.1880996}, which is further biased due to bias incurred during approximation.

Both the Ioffe's scheme and the biased un-weighted approximation scheme generate big hash values requiring 32-bits or higher storage per hash value. For reducing this to a manageable size of say 4-8 bits, a commonly adopted practical methodology is to randomly rehash it to smaller space at the cost of loss in accuracy~\cite{Article:Li_Konig_CACM11}. It turns out that our hashing scheme only generates 5-9 bits values, $h(x)$, satisfying Equation~\ref{eq:CollProb}, without losing any accuracy.

Next, we present a simple hashing procedure for weighted minwise hashing, which is unbiased and runs in expected constant time. We will later demonstrate that it can be around 60000x faster than Ioffe's scheme and around 100 times faster, in practice, than the fast un-weighted approximation.

\begin{algorithm}[h!]
\caption{Reduce to Unweighted~\cite{Report:Haeupler_arXiv14}}
\label{alg:UnWted}
\begin{algorithmic}
\INPUT Vector $x$,

\vspace{0.1in}
\STATE $S = \phi$
\FOR{Iterate over $x_j$ s.t $x_j > 0$ }
\STATE $floorx_j = \lfloor x_j \rfloor$
\FOR{$i=1$ {\bfseries to} $floorx_j$}
\STATE $S = S \cup (i,j)$
\ENDFOR
\STATE $r = Uniform(0,1)$
\IF {$r \le x_j - floorx_j$}
\STATE $S = S \cup (floorx_j+1,j)$
\ENDIF
\ENDFOR
\STATE {\bf RETURN} S (unweighted set)
\end{algorithmic}
\end{algorithm}

\section{Our Proposal: New Hashing Scheme}

We first describe our procedure in details. We will later talk about the correctness of the scheme, and we will show that our proposed scheme can be implemented in expected constant time.

\subsection{Procedure}
We will denote the $i^{th}$ component of vector $x \in \mathcal{R}^D$ by $x_i$. Let $m_i$ be the upper bound on the value of component $x_i$ in the given dataset. We can always assume the $m_i$ to be an integer, otherwise we take the ceiling $\lceil m_i\rceil$ as our upper bound. Define \begin{equation}\sum_{k=1}^i m_i = M_i. \ \ \ \text{and} \ \ \ \sum_{k=1}^D m_i = M_D = M\end{equation}
If the data is normalized, then $m_i = 1$ and $M = D$.

\begin{figure}[tb]
\vspace{-0.25in}
\label{fig:redgreen}
\includegraphics[width=3in,height=3in]{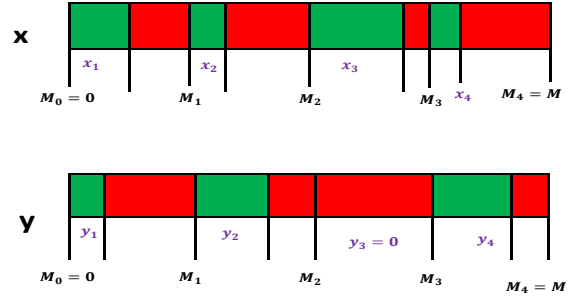}
\vspace{-1.2in}
\caption{Illustration of Red-Green Map of 4 dimensional vectors $x$ and $y$.}\vspace{-0.1in}
\end{figure}

Given a vector $x$, we first create a red-green map associated with it, as shown in Figure~\ref{fig:redgreen}. For this,
we first take an interval $[0,M]$ and divide it into $D$ disjoint intervals, with $i^{th}$ interval being $[M_{i-1},M_i]$ which is of the size $m_i$. Note that $\sum_{i=1}^D m_i = M$, so we can always do that. We then create two regions, red and green. For the $i^{th}$ interval $[M_{i-1},M_i]$, we mark the subinterval $[M_{i-1},M_{i-1} + x_i]$ as green and the rest $[M_{i-1} + x_i,M_i]$ with red, as shown in Figure~\ref{fig:redgreen}.  If $x_i =0$ for some $i$, then the whole $i^{th}$ interval $[M_{i-1},M_i]$ is marked as red.

Formally, for a given vector $x$, define the green $x_{green}$ and the red $x_{red}$ regions as follows
\begin{equation}\label{eq:redGreen}
  x_{green} = \cup_{i=1}^D [M_i,M_i+x_i]; \ \ \ \ x_{red} = \cup_{i=1}^D [M_i+x_i,M_{i+1}];
\end{equation}

\begin{algorithm}[h!]
\caption{Weighted MinHash}
\label{alg:HashPseudoAlgo}
\begin{algorithmic}
\INPUT Vector $x$, $M_i$'s, $k$, random seed[].

\vspace{0.1in}

\STATE Initialise Hashes[] to all 0s.
\FOR{$i=1$ {\bfseries to} $k$}
\STATE $randomseed = seed[i]$;
\WHILE{true}
\STATE $r = M \times Uniform(0,1)$;
\IF{ISGREEN(r), (check if $r \in {x_{red}}$}
\STATE break;
\ENDIF
\STATE $randomseed = \lceil r *1000000\rceil$ ;
\STATE $Hashes[i]++$;
\ENDWHILE
\ENDFOR
\STATE {\bf RETURN} Hashes
\end{algorithmic}
\end{algorithm}

Our sampling procedure simply draws an independent random real number between $[0,M]$, if the random number lies in the red region we repeat and re-sample. We stop the process as soon as the generated random number lies in the green region. Our hash value for a given data vector, $h(x)$, is simply the number of steps taken before we stop. We summarize the procedure in Algorithm~\ref{alg:HashPseudoAlgo}. More formally,
\begin{definition}
  Define $\{r_i : i = 1,2,3 ....\}$ as a sequence of i.i.d uniformly generated random number between $[0,M]$. Then we define the hash of $x$, $h(x)$ as
\begin{equation}
h(x) = \arg\min_{i} r_i, \ \ \ \text{s.t.} \ \ \  r_i \in x_{green}
\end{equation}
\end{definition}

We want our hashing scheme to be consistent~\cite{Proc:Ioffe_ICDM10} across different data points to guarantee Equation~\ref{eq:CollProb}. This requires ensuring the consistency of the random numbers in hashes~\cite{Proc:Ioffe_ICDM10}.
We can achieve the required consistency, either by pre-generating sequence of random numbers and storing them analogous to other hashing schemes. However, there is an easy way to generate a sequence of random numbers by ensuring the consistency of the random seed. This does not require any storage, except the starting seed. Our Algorithm~\ref{alg:HashPseudoAlgo} uses this criterion, to ensure that the sequence of random numbers is the same, we start with a fixed random seed for generating random numbers. If the generated random number lies in the red region, then before re-sampling, we reset the seed of our random number generator as a function of discarded random number. In the algorithm, we used $\lceil 100000*r \rceil$, where $\lceil \rceil$ is the ceiling operation, as a convenient way to ensure the consistency of sequence, without any memory overhead. This also works nicely in practice. Since we are sampling real numbers, the probability of any repetition (or cycle) is zero. For generating $k$ independent hashes we just use different random seeds which are kept fixed for the entire dataset.

\subsection{Proof of Correctness}

We show that the simple scheme given in Algorithms~\ref{alg:HashPseudoAlgo} actually does possess the required property, i.e. for any pair of points $x$ and $y$ Equation~\ref{eq:CollProb} holds. Unlike the previous works on this line~\cite{Report:Manasse_00,Proc:Ioffe_ICDM10} which requires computing the exact distribution of associated quantities, the proof of our proposed scheme is elementary and can be derived from first principles. This is not surprising given the simplicity of our procedure.

\begin{theorem}
\label{theo:main}
For any two vectors $x$ and $y$,  we have
\begin{equation}
Pr\bigg(h(x) = h(y)\bigg) =  \mathbb{J}(x,y) = \frac{\sum_{i=1}^D\min\{x_i,y_i\}}{\sum_{i=1}^D\max\{x_i,y_i\}}
\end{equation}
\end{theorem}
{\bf Proof:} First note that, every number between $[0,M]$ is random and equally likely in a random sampling. Therefore, for a given point $x$, at the time we stop we sample uniformly from the green region $x_{green} = \cup_{i=1}^D [M_i,M_i+x_i]$.  Consider the index $j$ defined as,
\begin{equation}
j = \min\{h(x),h(y)\}
\end{equation}
For any pair of points $x$ and $y$, consider the following three events: 1) $h(x) = h(y) =j$, 2) $h(x) > h(y) =j$ and 3) $j = h(x) < h(y)$.
Observe that,
\begin{align}
h(x) = h(y) =j  & \ \  \text{if and only if} \ \ r_j \in x_{green} \cap y_{green}\\
h(x) > h(y) =j  & \ \ \text{if and only if} \ \ r_j \in y_{green} - x_{green}\\
h(y) > h(x) =j  & \ \ \text{if and only if} \ \ r_j \in x_{green} - y_{green}
\end{align}
Since $r_j$ is uniformly chosen, we have,
\begin{align}\notag
&Pr\bigg(h(x) = h(y)\bigg) \\
\notag&= \frac{|x_{green} \cap y_{green}|}{|(x_{green} \cap y_{green}) \cup (x_{green} - y_{green}) \cup (y_{green} - x_{green})|}\\
&=\frac{|x_{green} \cap y_{green}|}{|x_{green} \cup y_{green}|}
\end{align}
The proof follows from substituting the values of $|x_{green} \cap y_{green}|$ and $|x_{green} \cup  y_{green}|$ given by:
\begin{align}\notag
|&x_{green} \cap  y_{green}| = |\cup_{i=1}^D [M_i,M_i+x_i] \cap \cup_{i=1}^D [M_i,M_i+y_i]| \\
&=|\cup_{i=1}^D [M_i,M_i+\min\{x_i,y_i\}]| =\sum_{i=1}^D\min\{x_i,y_i\}
\end{align}
and
\begin{align}\notag
|&x_{green} \cup  y_{green}| = |\cup_{i=1}^D [M_i,M_i+x_i] \cup \cup_{i=1}^D [M_i,M_i+y_i]| \\
&=|\cup_{i=1}^D [M_i,M_i+\max\{x_i,y_i\}]| =\sum_{i=1}^D\max\{x_i,y_i\},
\end{align}
Theorem~\ref{theo:main} implies that the sampling process is exact and we automatically have an unbiased estimator of $\mathbb{J}(x,y)$, using $k$ independently generated WMH, $h_i(x)$s from Algorithm~\ref{alg:HashPseudoAlgo}.
\begin{align}\label{eq:Estimator}
  \hat{J} = \frac{1}{k} \sum_{i=1}^k\big[{\bf 1}\{h_i(x) =& h_i(y)\}\big]; \ \ \mathbb{E}(\hat{J}) = \mathbb{J}(x,y);\\
   \ \  Var(\hat{J}) &= \frac{\mathbb{J}(x,y)(1 -\mathbb{J}(x,y))}{k},
\end{align}
where ${\bf 1}$ is the indicator function.

\subsection{Running Time Analysis and Fast Implementation}

Define
\begin{equation}\label{eq:sparsity}
  s_x = \frac{\text{Size of green region}}{\text{Size of red region}} = \frac{\sum_{i=1}^D x_i}{M}= \frac{||x||_1}{M},
\end{equation}
as the effective sparsity of the vector $x$. Note that this is also the probability of $Pr(r \in x_{green})$. Algorithm~\ref{alg:HashPseudoAlgo} has a while loop.

We show that the expected times the while loops runs, which is also the expected value of $h(x)$, is a small constant.
Formally,
\begin{theorem}
\label{theo:runtime}
\begin{align}\label{eq:Expec}
  \mathbb{E}(h(x)) = \frac{1}{s_x}; & \ \ Var(h(x)) = \frac{1-s_x}{s_x^2};\\
 \label{eq:probbound} Pr\bigg(h(x) \ge & \frac{\log{\delta}}{\log{(1-s_x)}}\bigg) \le \delta.
\end{align}
\end{theorem}
{\bf Proof:} Equation~\ref{eq:Expec} follows immediately from the fact that the number of sampling step taken before the process stops,which is also $h(x)$ is a geometric random variable with $p = s_x$. Equation~\ref{eq:probbound} follows from observing that $Pr(h(x) > k) = (1 - s_x)^k \le \delta$ which implies $k \le \frac{\log{\delta}}{\log{(1-s_x)}}$ yielding the required bound.

{\bf Remarks:} The time to compute each hash value in expectation is the inverse of effective sparsity $\frac{1}{s}$. Therefore, for datasets with $\frac{1}{s} << d$, we can expect our method to be much faster. For real datasets, such as image histograms, where minwise sampling is popular\cite{Proc:Ioffe_ICDM10}, the value of this sparsity is of the order of 0.02-0.08 (see Section~\ref{sec:experiments}) leading to $\frac{1}{s_x} \approx 13-50$. On the other hand, the number of non-zeros is around half million. Therefore, we can expect significant speed-up in practice.

As a consequence of Theorem~\ref{theo:runtime} we have:
\begin{corollary}
The expected amount of bits required to represent $h(x)$ is small, in particular,
\begin{align}\label{eq:ExpecBits}
  \mathbb{E}(bits) \le -\log{s_x}; & \ \ \mathbb{E}(bits) \approx \log{\frac{1}{s_x}} - \frac{(1-s_x)}{2};
\end{align}
\end{corollary}
{\bf Proof:} The proof follows from Jensens Inequality, $\mathbb{E}(\log{x}) \le \log{\mathbb{E}(x)}$
and second order Taylor series approximation of $\mathbb{E}(\log{x}) \approx \log{\mathbb{E}(x)} - \frac{Var(x)}{2 \log{\mathbb{E}(x)}^2}$ 

{\bf Remarks:} Existing hashing scheme require 64 bits, which is quite expensive. A popular approach for reducing space uses least significant bits of hashes~\cite{Article:Li_Konig_CACM11,Proc:Ioffe_ICDM10}. This tradeoff in space comes at the cost of accuracy~\cite{Article:Li_Konig_CACM11}. Our hashing scheme naturally requires only few bits, typically 5-9 (see Section~\ref{sec:experiments}), eliminating the need for trading accuracy for manageable space.

We know from Theorem~\ref{theo:runtime} that each hash function computation requires a constant number of function calls to ISGREEN(r). If we can implement ISGREEN(r) in constant time, i.e  $O(1)$, then we can  generate generate $k$ independent hashes in total $O(d + k)$ time instead of $O(dk)$ required by~\cite{Proc:Ioffe_ICDM10}. Note that $O(d)$ is the time to read the input vector which cannot be avoided. Once the data is loaded into the memory,  our procedure is actually O(k) for computing $k$ hashes, for all $k \ge 1$.

Before we jump into a constant time implementation, we would like readers to note that there is a straightforward binary search algorithm for ISGREEN(r) in $\log{d}$ time.  We consider $d$ intervals $[M_i,M_i+x_i]$ for all $i$, such that $x_i \ne 0$. Because of the nature of the problem, $M_{i-1} +x_{i-1} \le M_i$ $\forall i$. Therefore, these intervals are disjoint and sorted. Therefore, given a random number $r$, determining if $r \in \cup_{i=1}^D[M_i,M_i+x_i]$ only needs binary search over $d$ ranges. Thus, given that we only need constant number of random sampling in expectation before we stop, we already have a scheme that generates $k$ independent hashes in total $O(d + k\log{d})$ time improving over best known $O(dk)$ required by~\cite{Proc:Ioffe_ICDM10} for unbiased sampling.

We show that with some algorithmic tricks and few more data structures, we can implement ISGREEN(r) in constant time $O(1)$. We need two global pre-computed hashmaps, \emph{IntToComp} (Integer to Vector Component) and \emph{CompToM} (Vector Component to M value). $\emph{IntToComp}$ is a hashmap that maps every integer between $[0,M]$ to the associated components, i.e., all integers between $[M_i,M_{i+1}]$ are mapped to $i$, because it is associated with $i^{th}$ component. \emph{CompToM} maps every component of vectors $i \in \{1,\ 2,\ 3,...,\ D\}$ to its associated value $M_i$. The procedure for computing these hashmaps is straightforward and is summarized in Algorithm~\ref{alg:hashMaps}. It should be noted that these hash-maps computation is a one time pre-processing operation over the entire dataset having a negligible cost. $M_i$'s
can be computed (estimated) while reading the data.
\begin{algorithm}[tb]
\caption{ComputeHashMaps (Once per dataset)}
\label{alg:hashMaps}
\begin{algorithmic}
\INPUT $M_i$'s,
\STATE index =0, CompToM[0] =0
\FOR{$i=0$ {\bfseries to} $D-1$}
\IF{$i < D -1$}
\STATE $CompToM[i+1] = M_i + CompToM[i]$
\ENDIF
\FOR{$j=0$ {\bfseries to} $M_i -1$}
\STATE $IntToComp[index] = i$
\STATE index++
\ENDFOR
\ENDFOR
\STATE {\bf RETURN CompToM[] and IntToComp[]}
\end{algorithmic}
\end{algorithm}

\begin{algorithm}[h!]
\caption{ISGREEN(r)}
\label{alg:isgreen}
\begin{algorithmic}
\INPUT $r$, $x$, Hashmaps \emph{IntToComp[]} and \emph{CompToM[]} from Algorithm~\ref{alg:hashMaps}.

\vspace{0.1in}

\STATE $index = \lceil r \rceil$
\STATE $i = IntToComp[index]$
\STATE $M_i = CompToM[i]$
\IF{$r \le M_i + x_i$}
\STATE {\bf RETURN TRUE}
\ENDIF
\STATE {\bf RETURN FALSE}
\end{algorithmic}
\end{algorithm}

Using these two pre-computed hashmaps, the ISGREEN(r) methodology works as follows: We first compute the ceiling of $r$, i.e. $\lceil r\rceil$, then we find the component $i$  associated with $r$, i.e., $r \in [M_i,M_{i+1}]$, and the corresponding associated $M_i$ using hashmaps \emph{IntToComp} and \emph{CompToM}. Finally, we return true if $r \le x_i + M_i$ otherwise we return false. The main observation is that since we ensure that all $M_i$'s are Integers, for any real number r, if $r \in [M_i,M_{i+1}]$ then the same holds for $\lceil r\rceil$, i.e., $\lceil r\rceil \in [M_i,M_{i+1}]$. Hence we can work with hashmaps using $\lceil r\rceil$ as the key. The overall procedure of ISGREEN(r) is summarized in Algorithm~\ref{alg:isgreen}.

Note that our overall procedure is much simpler compared to Algorithm~\ref{alg:Ioffe}. We only need to generate random numbers followed by a simple condition check using two hash lookups. Our analysis shows that we have to repeat this only for a constant number of times. Compare this with the scheme of Ioffe where for every non-zero component of a vector we need to sample two Gamma variables followed by computing several expensive transformations including exponentials. We next demonstrate the benefits of our approach in practice.

\section{Practical Issues}

Theorem~\ref{theo:runtime} shows that the quantity $s_x = \frac{\sum_{i=1}^D x_i}{\sum_{i=1}^D M_i}$ determines the runtime. If $s_x$ is very very small then, although the running time is constant (independent of $d$ or $D$), it can still make the algorithm unnecessarily slow. Note that for the algorithm to work we choose $M_i$ to be the largest integer greater than the maximum possible value of co-ordinate $i$ in the given dataset. If this integer gap is big then we unnecessarily increase the running time. Ideally, the best running time is obtained when the maximum value, is itself an integer, or is very close to its ceiling value. If all the values are integers, scaling up does not matter, as it does not change $s_x$, but scaling down can make $s_x$ worse. Ideally we should scale, such that,  $\alpha = \arg\max_{\alpha} = \frac{\sum_{i=1}^D \alpha x_i}{\sum_{i=1}^D \lceil \alpha m_i \rceil}$ is  maximized, where $m_i$ is the maximum value of co-ordinate $i$ in  the given dataset.

\section{Experiments}
\begin{table}[ht]
\vspace{-0.15in}
\label{tab:stat}
\centering
\caption{Basic Statistics of the Datasets Used in the Paper}
\begin{tabular}{|p{2cm}|c|c|p{1.1cm}|} \hline
Data  & non-zeros (d) & Dim (D) & Sparsity (s) \\\hline
Web-Images Hist& 737 & 768& 0.081 \\\hline
Caltech101 & 95029 & 485640 & 0.024 \\\hline
Oxford & 401879 & 580644 & 0.086 \\
\hline\end{tabular}
\vspace{-0.1in}
\end{table}

We have proposed a new Weighted Minwise Hashing scheme which is significantly faster in theory than existing methodologies. In this section, we would like to assess its impact in practice. We compare the proposed scheme with the existing state-of-art-methodologies on estimation accuracy, running time and space. In particular, we will demonstrate that in real high-dimensional settings, our proposal provides significant speedup and requires less memory over existing methods. We also need to validate our theory that our scheme is unbiased and should be indistinguishable in accuracy with Ioffe's method.

{\bf Baselines:} Ioffe's method is the fastest known exact method in the literature, so it serves as our natural baseline. We also compare our method with biased unweighted approximations (see Algorithm~\ref{alg:UnWted}) which capitalizes on recent success in fast unweighted minwise hashing~\cite{Proc:OneHashLSH_ICML14,Proc:Shrivastava_UAI14}, we call it Fast-WDOPH (for Fast Weighted Densified One Permutation Hashing). Fast-WDOPH needs very long permutation, which is expensive. For efficiency, we implemented the permutation using fast $2$-universal hashing which is also what is recommended in practice~\cite{Proc:Mitzenmacher_08simple}.

{\bf Datasets:} Weighted Minwise sampling is commonly used for sketching image histograms~\cite{Proc:Ioffe_ICDM10}. We chose two popular publicly available vision dataset {\bf Caltech101}~\cite{Article:Fei_07CVIU} and {\bf Oxford}~\cite{Proc:Philbin_07CVPR}. We used the standard publicly available Histogram of Oriented Gradient (HOG) codes~\cite{Proc:Dalal_05CVPR}, popular in vision task, to convert images into feature vectors. In addition, we also used random web images ~\cite{Article:Wang_99} and computed simple histograms of RGB values. We call this dataset as {\bf Web-Images Hist}. The basic statistics of these datasets is summarized in Table~\ref{tab:stat}. These datasets cover a wide range of variations in terms of dimensionality, non-zeros and sparsity.

\begin{table}[ht]
\vspace{-0.15in}
\centering
\caption{Time taken in milliseconds (ms) to compute 500 hashes by different schemes. Our proposed scheme is significantly faster.}
\begin{tabular}{|c|c|c|p{1.2cm}|} \hline
Method & Prop &  Ioffe & Fast-WDOPH \\\hline
Web-Images Hist & 10ms & 986ms & 57ms \\\hline
Caltech101  & 57ms & 87105ms & 268ms  \\\hline
Oxford & 11ms &746120ms & 959ms \\
\hline\end{tabular}
\label{tab:time}
\vspace{-0.1in}
\end{table}

\subsection{Comparing Estimation Accuracy}

\begin{figure*}[ht]
\vspace{-0.1in}
\begin{center}

\mbox{
\includegraphics[width = 1.8in]{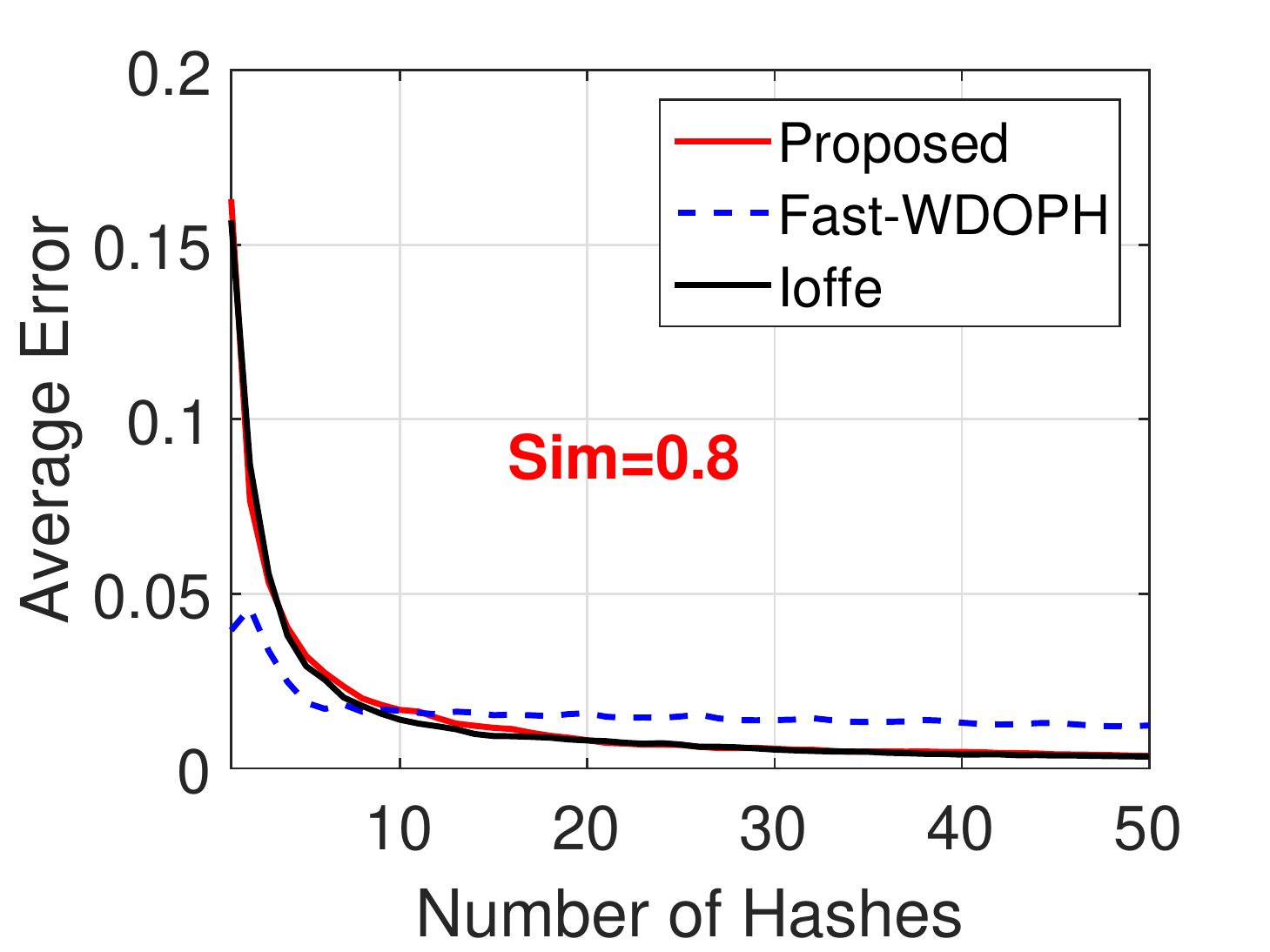} \hspace{-0.15in}
\includegraphics[width = 1.8in]{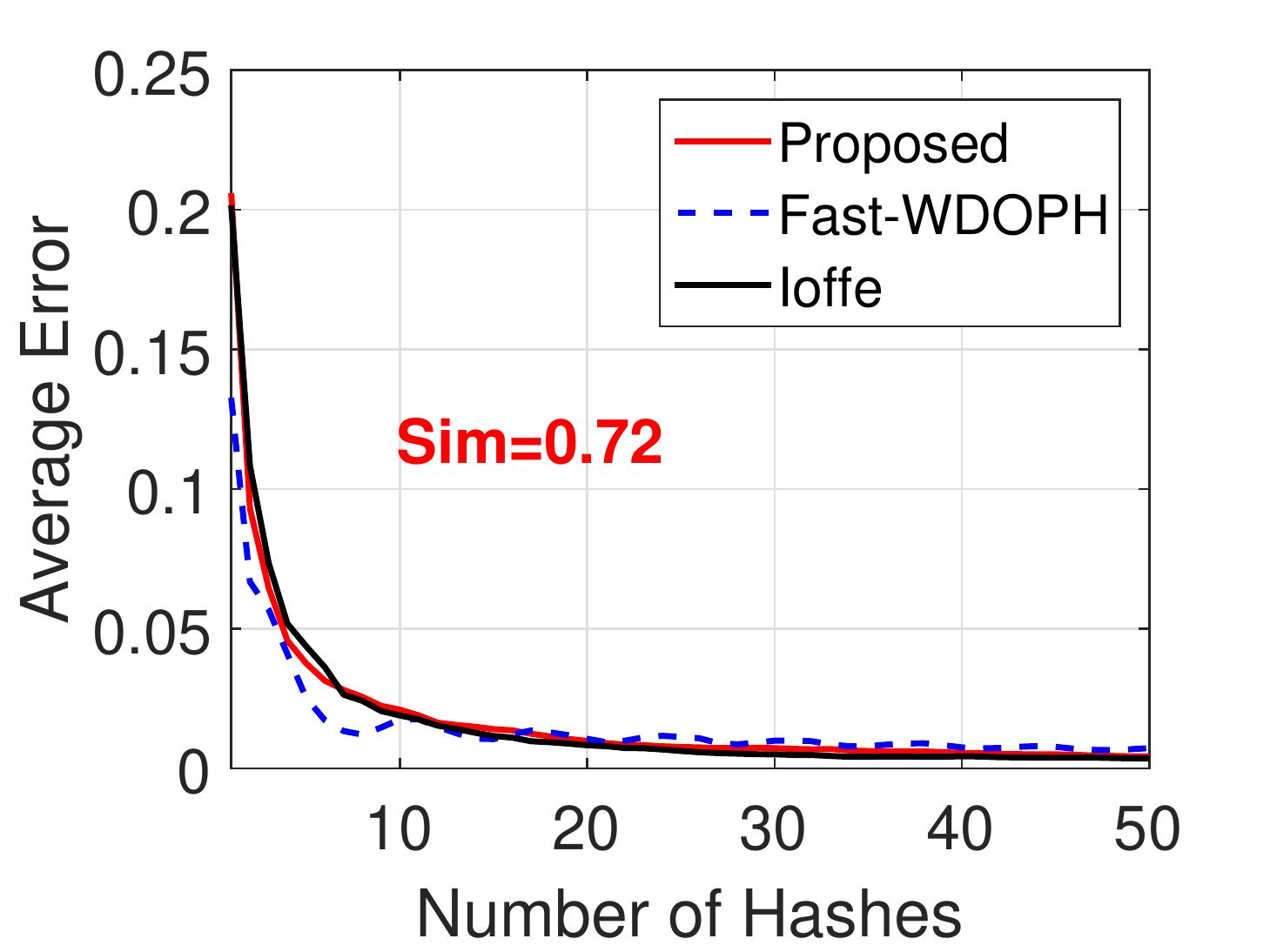}\hspace{-0.15in}
\includegraphics[width = 1.8in]{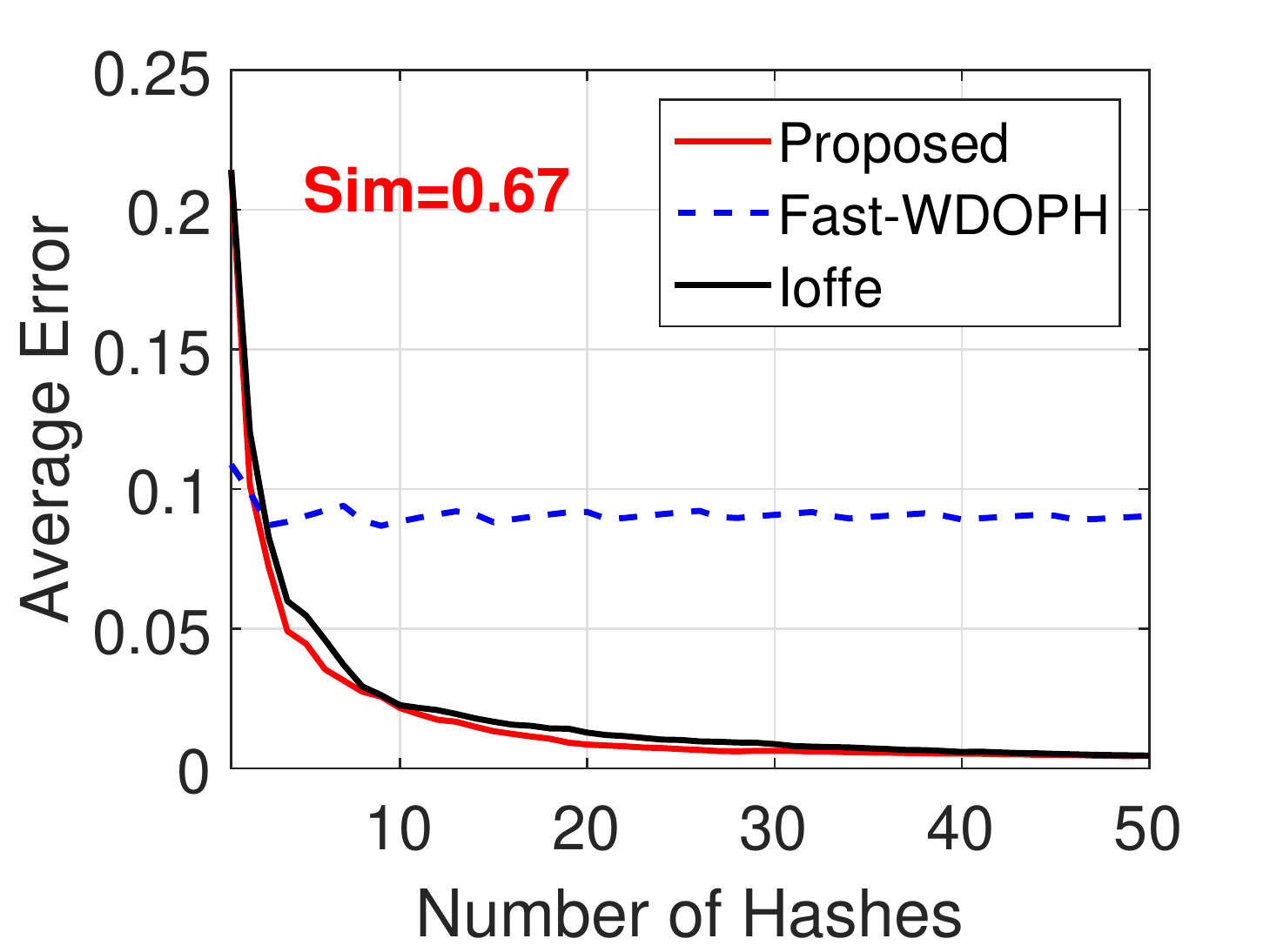}\hspace{-0.15in}
\includegraphics[width = 1.8in]{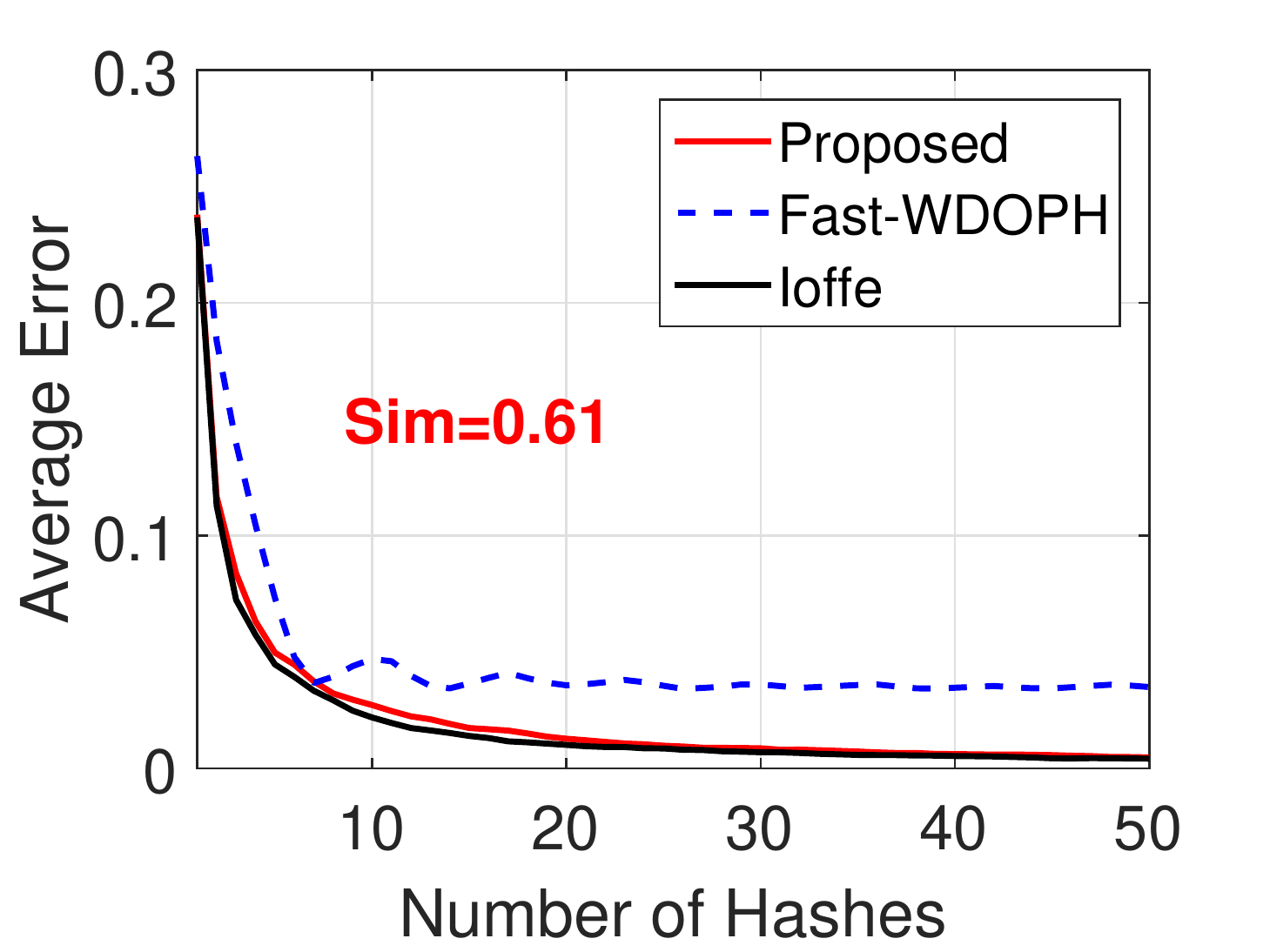}
}
\mbox{
\includegraphics[width = 1.8in]{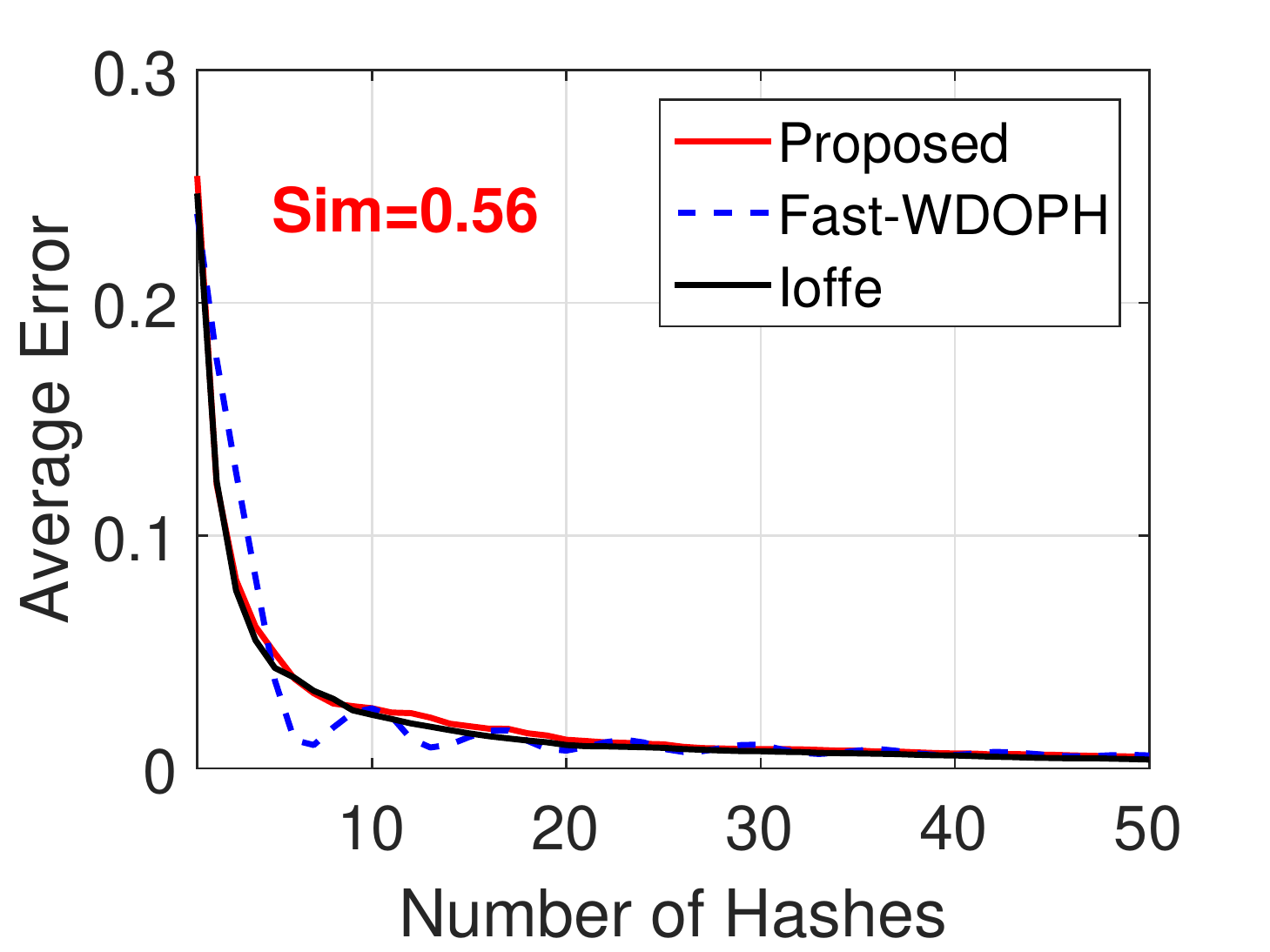}\hspace{-0.15in}
\includegraphics[width = 1.8in]{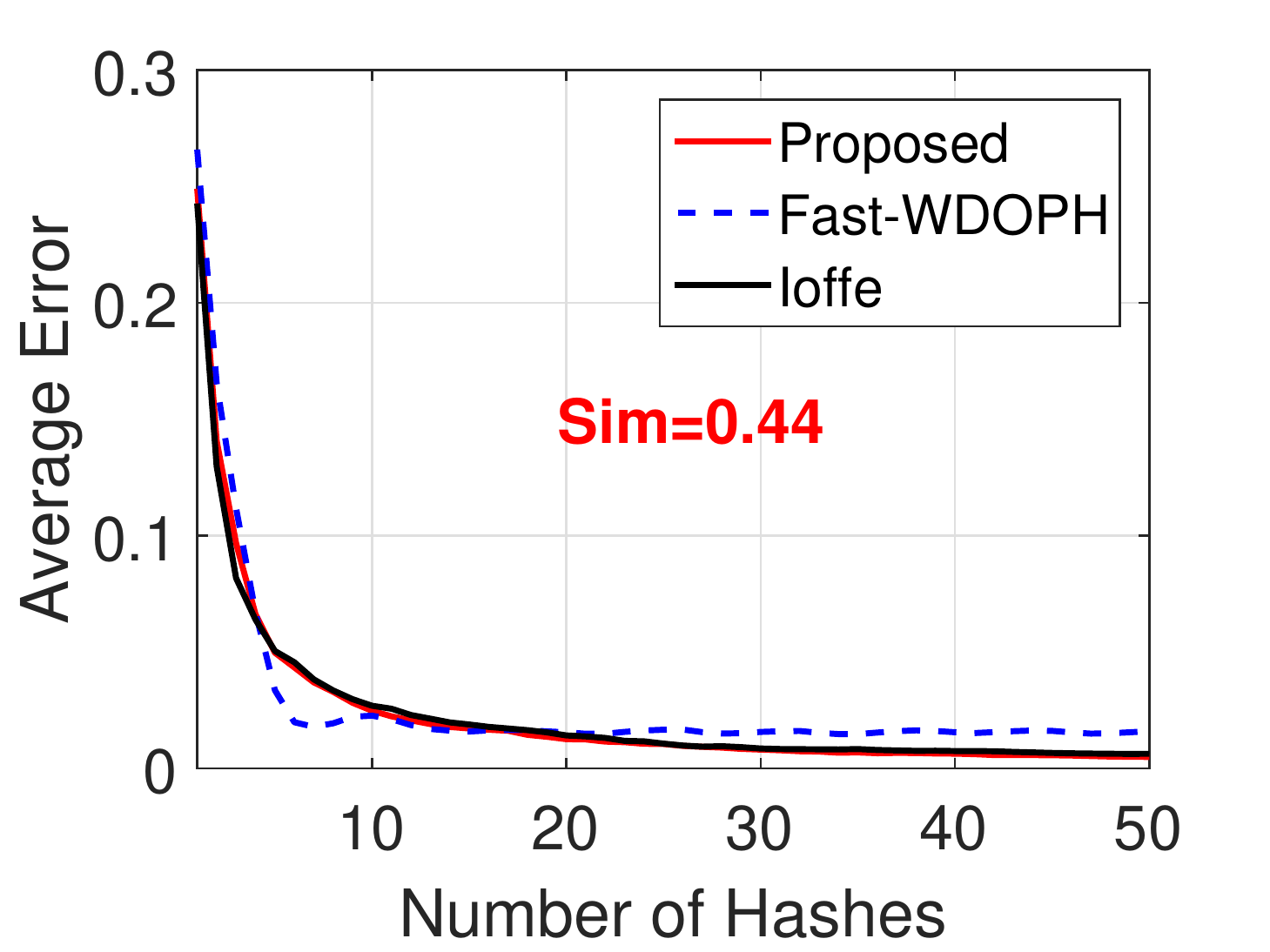}\hspace{-0.15in}
\includegraphics[width = 1.8in]{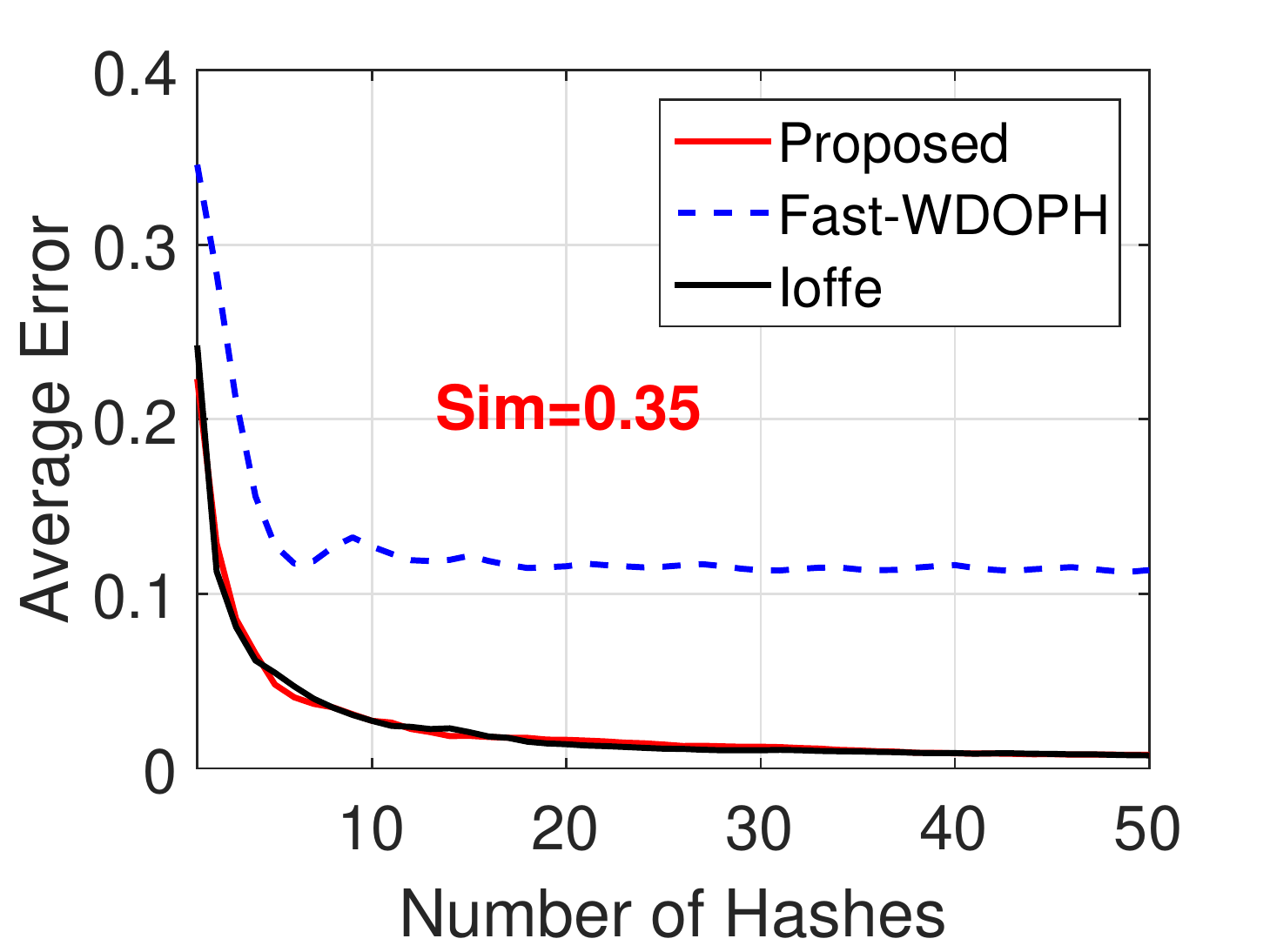}\hspace{-0.15in}
\includegraphics[width = 1.8in]{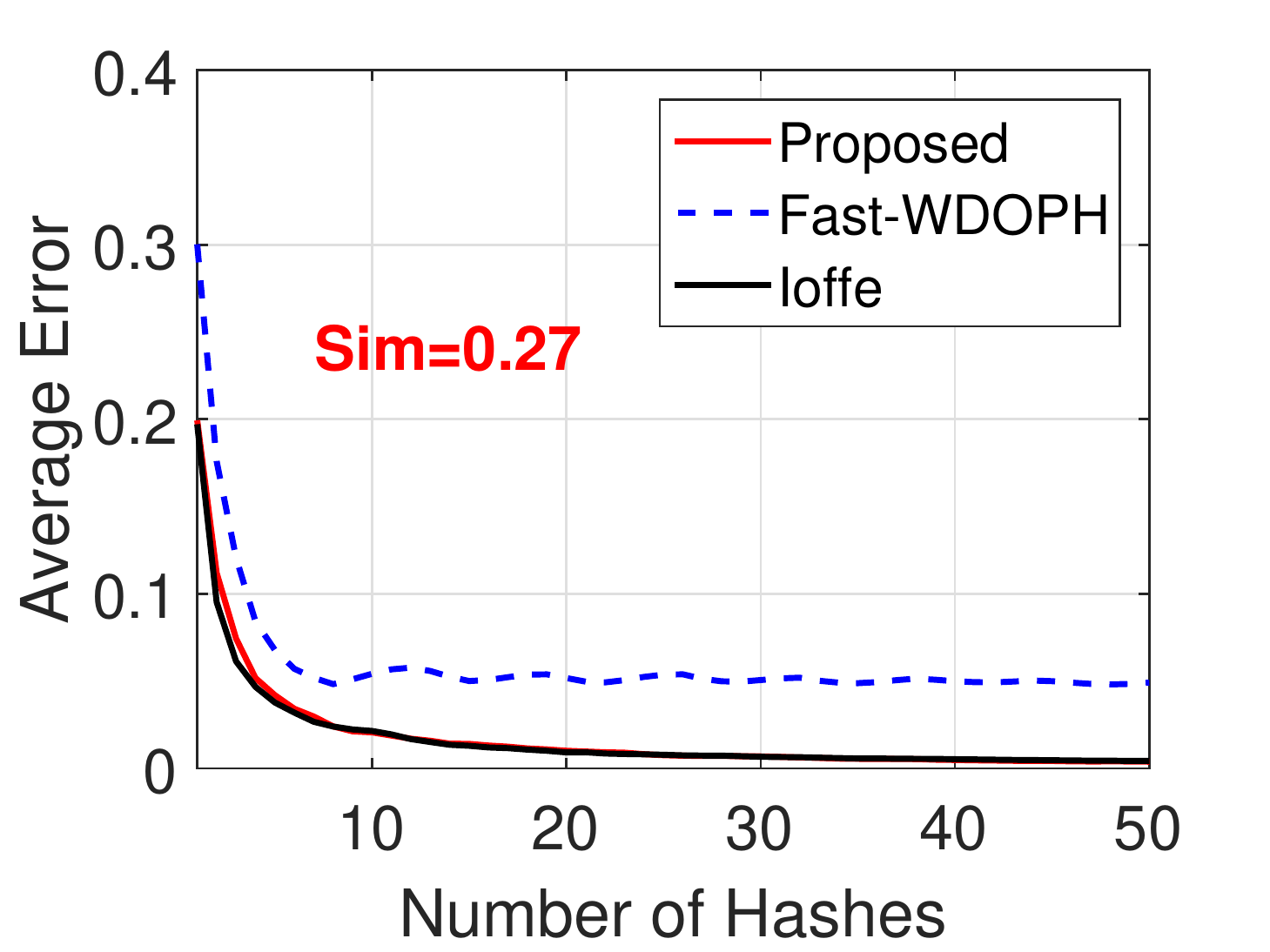}
}
\end{center}
\vspace{-0.1in}
\caption{\textbf{Average Errors in Jaccard Similarity Estimation with the Number of Hash Values. Estimates  are averaged over 200 repetitions.}}\label{fig:Errors}\vspace{-0.1in}
\end{figure*}

\begin{table}[h!]
\vspace{-0.19in}
\centering
\caption{The range of the observed hash values, using the proposed scheme, along with the maximum bits needed per hash value. The mean hash values agrees with Theorem~\ref{theo:runtime}}
\begin{tabular}{|p{2cm}|p{1.5cm}|p{1.5cm}|p{1.5cm}|} \hline
 &Web-Images Hist &  Caltech101  & Oxford \\\hline
  Mean Values & 11.94  &  52.88 & 9.13 \\\hline
 Hash Range &[1,107] & [1,487] &[1,69]  \\\hline
 Bits Needed &  7 & 9 & 7 \\
\hline
\end{tabular}
\label{tab:space}
\vspace{-0.1in}
\end{table}

In this section, we perform a sanity check experiment and compare the estimation accuracy with WMH. For this task we take 9 pairs of vectors from our datasets with varying level of similarities. For each of the pair $(x,y)$, we generate $k$ weighted minwise hashes $h_i(x)$ and $h_i(y)$ for $i \in \{1,\ 2,..,\ k\}$, using the three competing schemes. We then compute the estimate of the Jaccard similarity $\mathbb{J}(x,y)$ using the formula $\frac{1}{k} \sum_{i=1}^k\big[{\bf 1}\{h_i(x) = h_i(y)\}\big]$ (See Equation~\ref{eq:Estimator}). We compute the errors in the estimate as a function of $k$. To minimize the effect of randomization, we average the errors from 200 random repetitions with different seeds. We plot this average error with $k =\{1, \ 2, ..., \ 50\}$ in Figure~\ref{fig:Errors} for different similarity levels.

We can clearly see from the plots that the accuracy of the proposed scheme is indistinguishable from Ioffe's scheme. This is not surprising because both the schemes are unbiased and have the same theoretical distribution. This validates Theorem~\ref{theo:main}

The accuracy of Fast-WDOPH is inferior to that of the other two unbiased schemes and sometimes its performance is poor. This is because the weighted to unweighted reduction is biased and approximate. The bias of this reduction depends on the vector pairs under consideration, which can be unpredictable.

\subsection{Speed Comparisons}
\label{sec:experiments}

In this section, we demonstrate the actual benefits of having a constant time and low memory approach over existing methodologies. For this task, we compute the average time (in milliseconds) taken by the competing algorithms to compute 500 hashes of a given data vector for all the three datasets. Our experiments were coded in C\# on Intel Xenon CPU with 256 GB RAM.  Table~\ref{tab:time} summarises the comparison. We do not include the data loading cost in these numbers and assume that the data is in the memory for all the three methodologies.

We can clearly see tremendous speedup over Ioffe's scheme. For Web-Images dataset with mere 768 non-zeros, our scheme is 100 times faster than Ioffe's scheme and around 5 times faster than Fast-WDOPH approximation. While on caltech101 and Oxford datasets, which are high dimensional and dense datasets, our scheme can be 1500x to 70000x faster than Ioffe's scheme, while it is around 5 to  100x times faster than Fast-WDOPH scheme. Dense datasets like Caltech101 and Oxford represent more realistic scenarios. These features are taken from real applications~\cite{Proc:Dalal_05CVPR} and such level of sparsity and dimensionality are more common in practice.

The results are not surprising because Ioffe's scheme is very slow $O(dk)$. Moreover, the constant are inside bigO is also large, because of complex transformations. Therefore, for datasets with high values of $d$ (non-zeros) this scheme is very slow. Similar phenomena were observed in~\cite{Proc:Ioffe_ICDM10}, that decreasing the non-zeros by ignoring non-frequent dimensions can be around 150 times faster. However, ignoring dimension looses accuracy.

\subsection{Memory Comparisons}

Table~\ref{tab:space} summarizes the range of the hash values and the maximum number of bits needed to encode these hash values without any bias. We can clearly see that the hash values, even for such high-dimensional datasets, only require 7-9 bits. This is a huge saving compared to existing hashing schemes which can require (32-64) bits~\cite{Article:Li_Konig_CACM11}. Note, Ioffe's scheme generates a pair of integers as hashes which has significant storage cost. Thus, our method leads to around 5-6 times savings in space and it eliminates the need of shrinking hashes to small bits with a loss in the estimation accuracy. The mean  values observed (Table~\ref{tab:space}) validate the formula in Theorem~\ref{theo:runtime}.

\section{Conclusions}

We show that the existing popular weighted minwise hashing scheme can be made simpler, significantly faster and memory efficient. Our theoretical and empirical  results show that these advantages come for free and the hashes are indistinguishable from the unbiased scheme of Ioffe.

Experiments on real datasets show up to 60000 fold speedup in computing 500 hashes compared to the state-of-the-art procedure. We believe that our scheme will replace existing implementations in big-data systems.


\bibliography{mybib_merged}

\begin{thebibliography}{33}
\providecommand{\natexlab}[1]{#1}
\providecommand{\url}[1]{\texttt{#1}}
\expandafter\ifx\csname urlstyle\endcsname\relax
  \providecommand{\doi}[1]{doi: #1}\else
  \providecommand{\doi}{doi: \begingroup \urlstyle{rm}\Url}\fi

\bibitem[Andoni \& Indyk(2006)Andoni and Indyk]{Proc:Andoni_FOCS06}
Andoni, Alexandr and Indyk, Piotr.
\newblock Near-optimal hashing algorithms for approximate nearest neighbor in
  high dimensions.
\newblock In \emph{FOCS}, pp.\  459--468, Berkeley, CA, 2006.

\bibitem[Bayardo et~al.(2007)Bayardo, Ma, and Srikant]{Proc:Bayardo_WWW07}
Bayardo, Roberto~J., Ma, Yiming, and Srikant, Ramakrishnan.
\newblock Scaling up all pairs similarity search.
\newblock In \emph{WWW}, pp.\  131--140, 2007.

\bibitem[Broder(1997)]{Proc:Broder}
Broder, Andrei~Z.
\newblock On the resemblance and containment of documents.
\newblock In \emph{the Compression and Complexity of Sequences}, pp.\  21--29,
  Positano, Italy, 1997.

\bibitem[Broder(1998)]{Proc:Broder_FUN98}
Broder, Andrei~Z.
\newblock Filtering near-duplicate documents.
\newblock In \emph{FUN}, Isola d'Elba, Italy, 1998.

\bibitem[Broder et~al.(1997)Broder, Glassman, Manasse, and
  Zweig]{Proc:Broder_WWW97}
Broder, Andrei~Z., Glassman, Steven~C., Manasse, Mark~S., and Zweig, Geoffrey.
\newblock Syntactic clustering of the web.
\newblock In \emph{WWW}, pp.\  1157 -- 1166, Santa Clara, CA, 1997.

\bibitem[Charikar(2002)]{Proc:Charikar}
Charikar, Moses~S.
\newblock Similarity estimation techniques from rounding algorithms.
\newblock In \emph{STOC}, pp.\  380--388, Montreal, Quebec, Canada, 2002.

\bibitem[Chien \& Immorlica(2005)Chien and Immorlica]{Proc:Chien_WWW05}
Chien, Steve and Immorlica, Nicole.
\newblock Semantic similarity between search engine queries using temporal
  correlation.
\newblock In \emph{WWW}, pp.\  2--11, 2005.

\bibitem[Dalal \& Triggs(2005)Dalal and Triggs]{Proc:Dalal_05CVPR}
Dalal, Navneet and Triggs, Bill.
\newblock Histograms of oriented gradients for human detection.
\newblock In \emph{Computer Vision and Pattern Recognition}, volume~1, pp.\
  886--893. IEEE, 2005.

\bibitem[Datar et~al.(2004)Datar, Immorlica, Indyk, and
  Mirrokn]{Proc:Datar_SCG04}
Datar, Mayur, Immorlica, Nicole, Indyk, Piotr, and Mirrokn, Vahab~S.
\newblock Locality-sensitive hashing scheme based on $p$-stable distributions.
\newblock In \emph{SCG}, pp.\  253 -- 262, Brooklyn, NY, 2004.

\bibitem[Dwork \& Roth()Dwork and Roth]{Artice:Dwork_14}
Dwork, Cynthia and Roth, Aaron.
\newblock The algorithmic foundations of differential privacy.

\bibitem[Fei-Fei et~al.(2007)Fei-Fei, Fergus, and Perona]{Article:Fei_07CVIU}
Fei-Fei, Li, Fergus, Rob, and Perona, Pietro.
\newblock Learning generative visual models from few training examples: An
  incremental bayesian approach tested on 101 object categories.
\newblock \emph{Computer Vision and Image Understanding}, 106\penalty0
  (1):\penalty0 59--70, 2007.

\bibitem[Gollapudi \& Panigrahy(2006)Gollapudi and
  Panigrahy]{Proc:Gollapudi_06CIKM}
Gollapudi, Sreenivas and Panigrahy, Rina.
\newblock Exploiting asymmetry in hierarchical topic extraction.
\newblock In \emph{Proceedings of the 15th ACM international conference on
  Information and knowledge management}, pp.\  475--482. ACM, 2006.

\bibitem[Haeupler et~al.(2014)Haeupler, Manasse, and
  Talwar]{Report:Haeupler_arXiv14}
Haeupler, Bernhard, Manasse, Mark, and Talwar, Kunal.
\newblock Consistent weighted sampling made fast, small, and easy.
\newblock Technical report, arXiv:1410.4266, 2014.

\bibitem[Henzinge(2004)]{Article:Henzinger04}
Henzinge, Monika~.R.
\newblock Algorithmic challenges in web search engines.
\newblock \emph{Internet Mathematics}, 1\penalty0 (1):\penalty0 115--123, 2004.

\bibitem[Henzinger(2006)]{Proc:Henzinger_06}
Henzinger, Monika.
\newblock Finding near-duplicate web pages: a large-scale evaluation of
  algorithms.
\newblock In \emph{Proceedings of the 29th annual international ACM SIGIR
  conference on Research and development in information retrieval}, pp.\
  284--291. ACM, 2006.

\bibitem[Indyk(2001)]{Article:Indyk2001}
Indyk, Piotr.
\newblock A small approximately min-wise independent family of hash functions.
\newblock \emph{Journal of Algorithms}, 38\penalty0 (1):\penalty0 84--90, 2001.

\bibitem[Indyk \& Motwani(1998)Indyk and Motwani]{Proc:Indyk_STOC98}
Indyk, Piotr and Motwani, Rajeev.
\newblock Approximate nearest neighbors: Towards removing the curse of
  dimensionality.
\newblock In \emph{STOC}, pp.\  604--613, Dallas, TX, 1998.

\bibitem[Ioffe(2010)]{Proc:Ioffe_ICDM10}
Ioffe, Sergey.
\newblock Improved consistent sampling, weighted minhash and \text{L1}
  sketching.
\newblock In \emph{ICDM}, pp.\  246--255, Sydney, AU, 2010.

\bibitem[Kleinberg \& Tardos(1999)Kleinberg and Tardos]{Proc:Kleinberg_FOCS99}
Kleinberg, Jon and Tardos, Eva.
\newblock Approximation algorithms for classification problems with pairwise
  relationships: Metric labeling and \text{Markov} random fields.
\newblock In \emph{FOCS}, pp.\  14--23, New York, 1999.

\bibitem[Koudas et~al.(2006)Koudas, Sarawagi, and
  Srivastava]{Proc:Koudas_SIGMOD06}
Koudas, Nick, Sarawagi, Sunita, and Srivastava, Divesh.
\newblock Record linkage: similarity measures and algorithms.
\newblock In \emph{Proceedings of the 2006 ACM SIGMOD international conference
  on Management of data}, pp.\  802--803. ACM, 2006.

\bibitem[Li(2015)]{Proc:Li_KDD15}
Li, Ping.
\newblock 0-bit consistent weighted sampling.
\newblock In \emph{KDD}, 2015.

\bibitem[Li \& K\"onig(2011)Li and K\"onig]{Article:Li_Konig_CACM11}
Li, Ping and K\"onig, Arnd~Christian.
\newblock Theory and applications b-bit minwise hashing.
\newblock \emph{Commun. ACM}, 2011.

\bibitem[Manasse et~al.(2010)Manasse, McSherry, and Talwar]{Report:Manasse_00}
Manasse, Mark, McSherry, Frank, and Talwar, Kunal.
\newblock Consistent weighted sampling.
\newblock Technical Report MSR-TR-2010-73, Microsoft Research, 2010.

\bibitem[Mitzenmacher \& Vadhan(2008)Mitzenmacher and
  Vadhan]{Proc:Mitzenmacher_08simple}
Mitzenmacher, Michael and Vadhan, Salil.
\newblock Why simple hash functions work: exploiting the entropy in a data
  stream.
\newblock In \emph{Proceedings of the nineteenth annual ACM-SIAM symposium on
  Discrete algorithms}, pp.\  746--755. Society for Industrial and Applied
  Mathematics, 2008.

\bibitem[Philbin et~al.(2007)Philbin, Chum, Isard, Sivic, and
  Zisserman]{Proc:Philbin_07CVPR}
Philbin, J., Chum, O., Isard, M., Sivic, J., and Zisserman, A.
\newblock Object retrieval with large vocabularies and fast spatial matching.
\newblock In \emph{Proceedings of the IEEE Conference on Computer Vision and
  Pattern Recognition}, 2007.

\bibitem[P\u{a}tra\c{s}cu \& Thorup(2010)P\u{a}tra\c{s}cu and
  Thorup]{Patrascu:2010:KRL:1880918.1880996}
P\u{a}tra\c{s}cu, Mihai and Thorup, Mikkel.
\newblock On the k-independence required by linear probing and minwise
  independence.
\newblock In \emph{ICALP}, pp.\  715--726, 2010.

\bibitem[Rahimi \& Recht(2007)Rahimi and Recht]{Proc:Rahimi_NIPS07}
Rahimi, Ali and Recht, Benjamin.
\newblock Random features for large-scale kernel machines.
\newblock In \emph{Advances in neural information processing systems}, pp.\
  1177--1184, 2007.

\bibitem[Rajaraman \& Ullman()Rajaraman and Ullman]{Book:Raj_Ullman}
Rajaraman, Anand and Ullman, Jeffrey.
\newblock \emph{Mining of Massive Datasets}.

\bibitem[Rasheed \& Rangwala()Rasheed and Rangwala]{Proc:Rasheed_SIAM13}
Rasheed, Zeehasham and Rangwala, Huzefa.
\newblock Mc-minh: Metagenome clustering using minwise based hashing.
\newblock SIAM.

\bibitem[Shrivastava \& Li(2014{\natexlab{a}})Shrivastava and
  Li]{Proc:OneHashLSH_ICML14}
Shrivastava, Anshumali and Li, Ping.
\newblock Densifying one permutation hashing via rotation for fast near
  neighbor search.
\newblock In \emph{ICML}, Beijing, China, 2014{\natexlab{a}}.

\bibitem[Shrivastava \& Li(2014{\natexlab{b}})Shrivastava and
  Li]{Proc:Shrivastava_AISTATS14}
Shrivastava, Anshumali and Li, Ping.
\newblock $\{$In Defense of Minhash over Simhash$\}$.
\newblock In \emph{Proceedings of the Seventeenth International Conference on
  Artificial Intelligence and Statistics}, pp.\  886--894, 2014{\natexlab{b}}.

\bibitem[Shrivastava \& Li(2014{\natexlab{c}})Shrivastava and
  Li]{Proc:Shrivastava_UAI14}
Shrivastava, Anshumali and Li, Ping.
\newblock Improved densification of one permutation hashing.
\newblock In \emph{UAI}, Quebec, CA, 2014{\natexlab{c}}.

\bibitem[Wang et~al.(1999)Wang, Li, Chan, and Wiederhold]{Article:Wang_99}
Wang, J, Li, Jia, Chan, Desmond, and Wiederhold, Gio.
\newblock Semantics-sensitive retrieval for digital picture libraries.
\newblock \emph{D-Lib Magazine}, 5\penalty0 (11), 1999.

\end{thebibliography}
\bibliographystyle{icml2016}

\end{document}